\documentclass[aps,twocolumn]{revtex4}%
\usepackage{amsfonts}
\usepackage{amsmath}
\usepackage{amssymb}
\usepackage{appendix}
\usepackage{graphicx}%
\providecommand{\U}[1]{\protect\rule{.1in}{.1in}}
\providecommand{\U}[1]{\protect\rule{.1in}{.1in}}

\begin{document}
\title{Perfect optomechanically induced transparency in two-cavity optomechanics}

\author{\ Lai-Bin Qian}
\author{\ Xiao-Bo Yan}
\email{xiaoboyan@126.com}
\affiliation{School of Physics and Electronic Engineering, Northeast Petroleum University, Daqing, 163318, China}

\date{\today}

\begin{abstract}
Here, we study the controllable optical responses in a two-cavity optomechanical system, especially on the $\mathit{perfect}$ optomechanically induced transparency (OMIT) in the model which has never been studied before. The results show that the perfect OMIT can still occur even with a large mechanical damping rate, and at the perfect transparency window the long-lived slow light can be achieved. In addition, we find that the conversion between the perfect OMIT and optomechanically induced absorption can be easily achieved just by adjusting the driving field strength of the second cavity. We believe that the results can be used to control optical transmission in modern optical networks.
\end{abstract}
\maketitle

\section{Introduction}

The interaction between light and matter is an interesting and important research subject in quantum optics. It is an important means to understand the microstructure of matter.
Cavity optomechanics \cite{Aspelmeyer2014} can provide such a research platform where the macroscopic mechanical resonators and light fields interact with each other.   With the development of nanotechnology, various physical systems which can exhibit such interaction have been proposed and investigated, such as Fabry-Perot cavities \cite{Gigan2006,Arcizet2006}, whispering-gallery microcavities \cite{Kippenberg2005,Tomes2009,Jiang2009}, superconducting circuits \cite{Regal2008,Teufel2011_471} and membranes \cite{Thompson2008,Jayich2008,Sankey2010,Karuza2013}.
The optomechanical interaction can strongly affect the motion of mechanical oscillator and the optical properties in these systems, and then various interesting quantum phenomena can be generated, such as ground-state cooling of mechanical modes
\cite{Marquardt2007,Wilson-Rae2007,BingHe2017,Wang2018pra,Yang2019OE,Wang2021CPB,He2022pra,Yang2022FOP}, quantum entanglement
\cite{Yan2019OE,Li2017pra,Bai2016SR,Deng2016pra,Yan2017pra,Chen2021PRA,Guan2022njpQI,Zeng2022FOP}, mechanical squeezing \cite{Bai2019PR}, unconventional photon blockade \cite{Feng2023FOP}, and optomechanically induced transmission and absorption \cite{Yan2014,Qu2013pra,Agarwal2014njp,Wang2022QIP,He2016pra,Du2018epl,Xia2019oc,Yan2019FOP,Chen2020OE,Wang2021PRA,Qi2021FOP,Zhong2022FOP}.

Recently, the researches on optomechanically induced transparency (OMIT) \cite{Huang2010_041803,Weis2010,Safavi-Naeini2011,Xiong2012pra,Yan2020pra,Shahidani2013,LiuYX2013,Kronwald2013prl,Jing2015sr,Li2016sr,Liu2017Nano,Bodiya2019pra,Yan2021jpb,PanGx2021lp,Yan2014EPJD} and the associated slow light in optomechanical systems have attracted much attentions. A remarkable feature of OMIT is that there is a deep dip in the absorption curve accompanied with a steep dispersion behavior at the transparency window. The steep dispersion behavior can generate a drastic reduction in the group velocity of light passing through the system \cite{Safavi-Naeini2011}. According to this effect, many schemes have been proposed to slow or stop  light \cite{Chen2011,Chang2011njp,Tarhan2013,Akram2015,Gu2015,Safavi-Naeini2011,Yan2021pe,Yan2021jpb,Wanglan2020lp,Zhao2021photo}, which is very meaningful in the construction of quantum information networks.
However, the ideal depth of the transparency window cannot be achieved  due to the nonzero mechanical damping rate in the usual  OMIT theory \cite{Huang2010_041803,Safavi-Naeini2011,Weis2010},
which will cause a very limited slow light effect (usually on the order of milliseconds) \cite{Chen2011,Tarhan2013,Gu2015}.
Until recently, the perfect OMIT can be easily achieved using the mechanism of non-rotating wave approximation\cite{Yan2020pra,Yan2021jpb}, and the slow light effect  at the perfect transparency window can be greatly improved \cite{Yan2021jpb}. The perfect OMIT theory above is studied in standard single-cavity optomechanics. In fact, there are more abundant and  interesting quantum phenomena in multi-mode optomechanical systems, such as quantum entanglement
\cite{Yan2019OE,Deng2016pra,Yan2017pra}, optical nonreciprocity \cite{Xu2015pra,Bernier2017nc} and  quantum nonlinearity \cite{Ludwig2012prl,lv2013sr}.  Therefore, it is necessary to generalize the perfect OMIT theory to multi-mode optomechanical systems.

Here, we theoretically study the optical responses including perfect OMIT, long-lived slow light and optomechanically induced absorption in a multi-mode optomechanical system (comprising two optical modes and one mechanical mode), see Fig. 1 \cite{Yan2019OE,Ludwig2012prl,Tian2013prl,lv2013sr}. First, we give the conditions under which the perfect OMIT will occur, and the window width expression of the perfect OMIT. As long as the conditions are satisfied, the perfect OMIT can be achieved even with a large mechanical damping rate. 
Secondly, the dispersion curve becomes very steep at the perfect transparency window where it can be proofed that the negative dispersion curve slope is exactly equal to the value of time delay (slow light) in the model. It means that the long-lived slow light can be achieved  at the perfect transparency window.
Thirdly, the driving strength and dissipation of the second cavity have a great effect on the optical response of the system. Especially, the conversion between perfect OMIT and optomechanically induced absorption in the model can be easily achieved just by adjusting the driving field strength of the second cavity, which can be used as an optical switch in modern optical networks.

\section{Model and equations}

\begin{figure}[ptb]
	\includegraphics[width=0.45\textwidth]{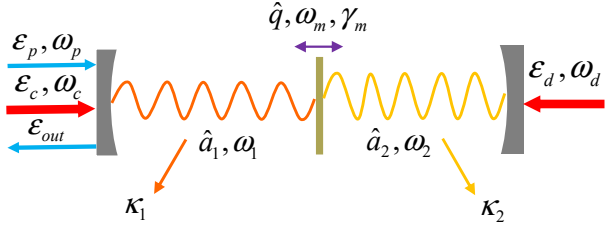}\caption{Sketch of a two-cavity optomechanical system consists of a mechanical membrane with frequency $\omega_{m}$ interacted with cavity $\hat{a}_{1}$ and $\hat{a}_{2}$ via radiation pressure. The cavity $\hat{a}_{1}$ is driven by a coupling field $\varepsilon_{c}$ with frequency $\omega_{c}$ and a weak probe field $\varepsilon_{p}$ with frequency $\omega_{p}$. The cavity $\hat{a}_{2}$ is driven by a driving field $\varepsilon_{d}$ with frequency $\omega_{d}$.}%
	\label{Fig1}%
\end{figure}

We consider an optomechanical system consisting of a mechanical membrane and two cavities (see Fig. 1). The mechanical membrane with frequency $\omega_{m}$ and mass $m$ is coupled to the two cavities via radiation pressure effects. The frequencies of the two cavities are described as $\omega_{1}$ and $\omega_{2}$, and the annihilation operators of the two cavities are denoted by $\hat{a}_{1}$ and $\hat{a}_{2}$, respectively. The position and momentum operators of the mechanical membrane are represented by $\hat{q}$ and $\hat{p}$, respectively.
The radiation pressure effect can be described by $\hbar\hat{q}(g_{2}\hat{a}_{2}^{\dagger}\hat{a}_{2} -g_{1}\hat{a}_{1}^{\dagger}\hat{a}_{1})$ with the optomechanical coupling rate $g_{i}=\omega_{i}/L_{i}$ where $L_{i}$ is the geometric length of the cavity $i$. The cavity $\hat{a}_{1}/\hat{a}_{2}$ is driven by a strong coupling/driving field with
frequency $\omega_{c}/\omega_{d}$ and amplitude $\varepsilon_{c}/\varepsilon_{d}$. In addition, a weak probe field with frequency $\omega_{p}$ and amplitude $\varepsilon_{p}$ is injected into cavity $\hat{a}_{1}$. For simplicity and without loss of generality, we can set $\omega_{1}=\omega_{2}=\omega_{0}$ and $L_{1}=L_{2}=L$. Then, the Hamiltonian of the system $H_{s}$ is 
\begin{eqnarray}
H_{s}=\hbar\omega_{0}\hat{a}_{1}^{\dagger }\hat{a}_{1}+\hbar\omega_{0}\hat{a}_{2}^{\dagger }\hat{a}_{2}+\frac{\hat{p}^{2}}{2m}+\frac{1}{2}m\omega_{m}^{2}\hat{q}^{2}\qquad\quad\\
+\hbar\hat{q}g_{0}(\hat{a}_{2}^{\dagger}\hat{a}_{2} -\hat{a}_{1}^{\dagger}\hat{a}_{1})+i\hbar(\hat{a}_{1}^{\dagger}\varepsilon_{p}e^{-i\omega_{p}t}-\hat{a}_{1}\varepsilon_{p}^{\ast}e^{i\omega_{p}t})\quad \notag\\
+i\hbar\varepsilon_{c}(\hat{a}_{1}^{\dagger}e^{-i\omega_{c}t}-\hat{a}_{1}e^{i\omega_{c}t})+i\hbar\varepsilon_{d}(\hat{a}_{2}^{\dagger}e^{-i\omega_{d}t}-\hat{a}_{2}e^{i\omega_{d}t})\notag  
\end{eqnarray}%
with $g_{0}=\omega_{0}/L$.

Since in quantum mechanics the unitary transformation will not change the physics properties of the system, we can take the unitary transformation $\hat{U}=e^{i(\omega_{c}\hat{a}_{1}^{\dagger }\hat{a}_{1}+\omega_{d}\hat{a}_{2}^{\dagger }\hat{a}_{2})t}$ using the formula $H=\hat{U}H_{s}\hat{U}^{\dagger}-i\hbar\hat{U}\frac{\partial \hat{U}^{\dagger}}{\partial t}$. After this transformation, the Hamiltonian $H_{s}$ will turn to $H$ which can be given by
\begin{eqnarray}
H&=&\hbar\Delta_{c}\hat{a}_{1}^{\dagger }\hat{a}_{1}+\hbar\Delta_{d}\hat{a}_{2}^{\dagger }\hat{a}_{2}+\frac{\hat{p}^{2}}{2m}+\frac{1}{2}m\omega_{m}^{2}\hat{q}^{2}\notag\\
&+&\hbar\hat{q}g_{0}(\hat{a}_{2}^{\dagger}\hat{a}_{2} -\hat{a}_{1}^{\dagger}\hat{a}_{1})+i\hbar(\hat{a}_{1}^{\dagger}\varepsilon_{p}e^{-i\delta t}-\hat{a}_{1}\varepsilon_{p}^{\ast}e^{i\delta t})\notag\\
&+&i\hbar\varepsilon_{c}(\hat{a}_{1}^{\dagger}-\hat{a}_{1})+i\hbar\varepsilon_{d}(\hat{a}_{2}^{\dagger}-\hat{a}_{2}).  
\end{eqnarray}%
Here,  $\Delta_{c/d}=\omega_{0}-\omega_{c/d}$ is the detuning between cavity field $\hat{a}_{1}/\hat{a}_{2}$ and coupling/driving field, and  $\delta=\omega_{p}-\omega_{c}$ is the detuning between the probe field and coupling field.

In this paper, we deal with the mean response of the system to the probe field in the presence of the coupling and driving field, hence we do not include quantum fluctuations.
According to the Heisenberg-Langevin equation and from Eq. (2), we can obtain the  motion equations of mean values of the system operators as follow
\begin{eqnarray}
\langle\dot{\hat{a}}_{1}\rangle&=&-[\kappa_{1}+i(\Delta_{c}-g_{0}\langle \hat{q}\rangle)]\langle \hat{a}_{1}\rangle+\varepsilon_{c}+\varepsilon_{p}e^{-i\delta t},\notag\\
\langle\dot{\hat{a}}_{2}\rangle&=&-[\kappa_{2}+i(\Delta_{d}+g_{0}\langle \hat{q}\rangle)]\langle \hat{a}_{2}\rangle+\varepsilon_{d},\notag\\
\langle\dot{\hat{p}}\rangle&=&-\gamma_{m}\langle \hat{p}\rangle-m\omega_{m}^{2}\langle \hat{q}\rangle+\hbar g_{0}(\langle \hat{a}_{1}^{\dagger}\rangle\langle \hat{a}_{1}\rangle-\langle \hat{a}_{2}^{\dagger}\rangle\langle \hat{a}_{2}\rangle),\notag\\
\langle\dot{\hat{q}}\rangle&=&\frac{\langle \hat{p}\rangle}{m}.
\end{eqnarray}
Here, we have used the usual factorization assumption, i.e. $\langle \hat{\mathrm{X}}\hat{\mathrm{Y}}\rangle=\langle\hat{\mathrm{X}}\rangle\langle\hat{\mathrm{Y}}\rangle$ (which holds in the case of single-photon
weak coupling, i.e., $g_{0}\ll\omega_{m}$ \cite{Nunnenkamp2011PRL}), and $\gamma_{m}$, $\kappa_{1}$ and $\kappa_{2}$ are the damping rates of the mechanical membrane,  cavity $\hat{a}_{1}$ and $\hat{a}_{2}$, respectively.

It is very difficult to obtain the exact solution of Eq. (3) because it is a nonlinear equation. However, in general the strength of the probe field is much smaller than that of the coupling field in cavity optomechanics. Hence, we can solve Eq. (3) by perturbation method. To this end, we assume that the solution of the mean values of the operators in Eq. (3) has the following form
\begin{eqnarray}
\langle \hat{s}\rangle=s_{0}+\varepsilon_{p}e^{-i\delta t}s_{+}+\varepsilon_{p}^{\ast}e^{i\delta t}s_{-}
\end{eqnarray}
where $s=\{q,\:p,\:a_{1},\:a_{2}\}$.
It means that we can safely ignore the higher order terms of the probe field $\varepsilon_{p}$.

We are particularly interested in the properties of the field with frequency $\omega_{p}$ in the output field of cavity $\hat{a}_{1}$, which can be determined by the term $a_{1+}$ in Eq. (4). The term $a_{1-}$ denotes the anti-Stokes effect and we don't care here.   
Substituting Eq. (4) in Eq. (3) and ignoring the higher order terms of $\varepsilon_{p}$, we can obtain the important expression for the term $a_{1+}$ (See Appendix A  for detailed calculations) as
\begin{widetext}
\begin{eqnarray}
a_{1+}=\frac{1}{\kappa_{1}-i(\delta-\Delta_{1})+\frac{\beta_{1}}{\frac{\delta^{2}-\omega_{m}^{2}+i\delta\gamma_{m}}{2i\omega_{m}}-\frac{\beta_{1}}{\kappa_{1}-i(\delta+\Delta_{1})}+\frac{\beta_{2}}{\kappa_{2}-i(\delta-\Delta_{2})}-\frac{\beta_{2}}{\kappa_{2}-i(\delta+\Delta_{2})}}},
\end{eqnarray}
\end{widetext}
here,
\begin{eqnarray}
\beta_{1}=\frac{\hbar g_{0}^{2}|a_{10}|^{2}}{2m\omega_{m}},\quad \beta_{2}=\frac{\hbar g_{0}^{2}|a_{20}|^{2}}{2m\omega_{m}},
\end{eqnarray}
and $\Delta_{1}=\Delta_{c}-g_{0}q_{0}$, 
$\Delta_{2}=\Delta_{d}+g_{0}q_{0}$, and the expressions of $q_{0}$, $a_{10}$ and $a_{20}$ can be found in Appendix A.

Because it is known that the coupling between the cavity and the resonator is strong at the near-resonant frequency, in this paper, we consider $\Delta_{1}\sim\Delta_{2}\sim\omega_{m}$ and $\delta\sim\omega_{m}$. Then we have $\delta^{2}-\omega_{m}^{2}\sim2\omega_{m}(\delta-\omega_{m})$ and $\delta+\Delta_{1}\sim\delta+\Delta_{2}\sim2\omega_{m}$. If we set $x=\delta-\omega_{m}$, Eq. (5) can be simplified as
\begin{eqnarray}
a_{1+}=\frac{1}{\kappa_{1}-ix+\frac{\beta_{1}}{\frac{\gamma_{m}}{2}-ix-\frac{\beta_{1}}{\kappa_{1}-2i\omega_{m}}+\frac{\beta_{2}}{\kappa_{2}-ix}-\frac{\beta_{2}}{\kappa_{2}-2i\omega_{m}}}}.
\end{eqnarray}
Next, based on Eq. (7), we will study the optical response of the optomechanical system to the probe field, including the $\mathit{perfect}$ optomechanically induced transparency, slow light and optomechanically induced absorption, respectively.

\section{Perfect optomechanically induced transparency}

According to input-output relation \cite{Huang2010_041803}, the quadrature of the  optical component with frequency $\omega_{p}$ in the output field
can be defined as $\varepsilon_{T}=2\kappa_{1}a_{1+}$ \cite{Huang2010_041803}. The real part $\mathrm{Re}[\varepsilon_{T}]$ and imaginary part $\mathrm{Im}[\varepsilon_{T}]$ represent the absorptive and dispersive behavior of the optomechanical system to the probe field, respectively. We first give the conditions of perfect OMIT in the model. According to the conclusions in Ref. \cite{Yan2020pra}, the conditions of perfect OMIT are determined by the pole location of the subfraction in Eq. (7), i.e., 
\begin{eqnarray}
\frac{\gamma_{m}}{2}-ix-\frac{\beta_{1}}{\kappa_{1}-2i\omega_{m}}+\frac{\beta_{2}}{\kappa_{2}-ix}-\frac{\beta_{2}}{\kappa_{2}-2i\omega_{m}}=0.
\end{eqnarray}
It can be known from Eq. (8) that the perfect OMIT cannot occur at the resonant frequency $x=0$ because in this case $\beta_{2}=-\frac{\gamma_{m}\kappa_{2}(\kappa_{2}^{2}+4\omega_{m}^{2})}{2(\kappa_{1}\kappa_{2}+4\omega_{m}^{2})}<0$ which is in contradiction with the definition in Eq. (6).

We first study the case of large $\kappa_{2}$, i.e., the cavity damping rate $\kappa_{2}$ is much larger than the detuning $x$ where the perfect OMIT appears ($\kappa_{2}\gg|x|$), and from Eq. (8) we can obtain the conditions of perfect OMIT as
\begin{eqnarray}
x=\frac{2\kappa_{2}\omega_{m}[\gamma_{m}\kappa_{2}(\kappa_{1}^{2}+4\omega_{m}^{2})-2\beta_{1}(\kappa_{1}\kappa_{2}+4\omega_{m}^{2})]}{(\kappa_{1}^{2}+4\omega_{m}^{2})\xi-2\beta_{1}\kappa_{1}(\kappa_{2}^{2}+4\omega_{m}^{2})},\notag\\
\beta_{2}=\frac{\kappa_{2}(\kappa_{2}^{2}+4\omega_{m}^{2})(2\beta_{1}\kappa_{1}-\gamma_{m}(\kappa_{1}^{2}+4\omega_{m}^{2}))}{8\omega_{m}^{2}(\kappa_{1}^{2}+4\omega_{m}^{2})},
\end{eqnarray}
with $\xi=8\kappa_{2}\omega_{m}^{2}+\gamma_{m}(\kappa_{2}^{2}+4\omega_{m}^{2})$.

\begin{figure}[t]
	\includegraphics[width=0.47\textwidth]{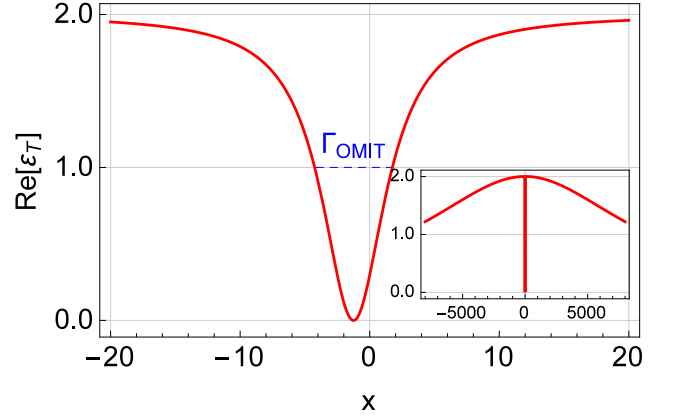}\caption{The real part $\mathrm{Re}[\varepsilon_{T}]$ (red-solid) vs. frequency detuning $x$ with parameters $\omega_{m}=\kappa_{1}=\kappa_{2}=10^{4}$, $\gamma_{m}=1$, $\beta_{1}=3\times10^{4}$ and $\beta_{2}=1250$ according to Eq. (9). The blue-dashed line indicates width $\Gamma_{\mathrm{OMIT}}$. The inset in Fig. 2 shows the OMIT profile in a large scale.}%
	\label{Fig2}%
\end{figure}

\begin{figure}[ptb]
	\includegraphics[width=0.46\textwidth]{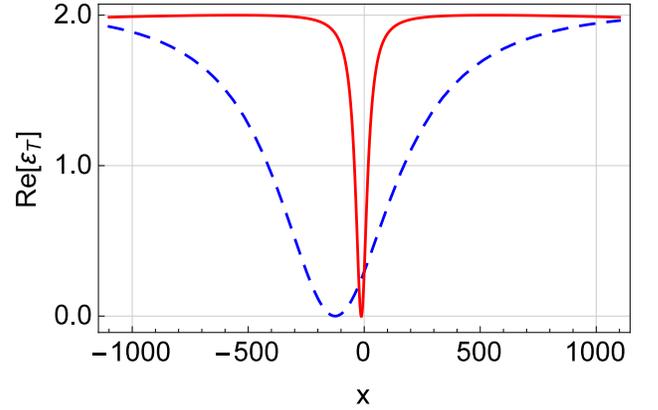}\caption{The  real part  $\mathrm{Re}[\varepsilon_{T}]$ vs. frequency detuning $x$ for large mechanical damping rate $\gamma_{m}=10$ (red-solid) with $\beta_{1}=3\times10^{5}$ and $\beta_{2}=1.25\times10^{4}$ according to Eq. (9), and for $\gamma_{m}=100$ (blue-dashed) with $\beta_{1}=3\times10^{6}$ and $\beta_{2}=1.25\times10^{5}$ according to Eq. (9). The other parameters are same as Fig. 2.}%
	\label{Fig3}%
\end{figure}

The perfect OMIT can be achieved if the conditions in Eq. (9) are satisfied. In Fig. (2), we plot the real part $\mathrm{Re}[\varepsilon_{T}]$ (red-solid) vs. frequency detuning $x$ with $\omega_{m}=\kappa_{1}=\kappa_{2}=10^{4}$, $\gamma_{m}=1$, $\beta_{1}=3\times10^{4}$ and $\beta_{2}=1250$ according to Eq. (9). With these parameters, according to Eq. (9), the transparency window will appear at $x\simeq-1.25$ which is consistent with the numerical result in Fig. (2). The inset in Fig. (2) shows the OMIT profile in a large scale, from which it can be seen that the width of the transparency window is very narrow. From Eq. (7), we can obtain the expression of the width $\Gamma_{\mathrm{OMIT}}$ (full width at half maximum) of transparent window, but it is too lengthy to be reported here. However, if we take the case of equal cavity damping rate ($\kappa_{1}=\kappa_{2}=\kappa$), the width $\Gamma_{\mathrm{OMIT}}$ can be given as
\begin{eqnarray}
\Gamma_{\mathrm{OMIT}}=\frac{\sqrt{\kappa[32\beta_{1}\omega_{m}^{2}\eta+\kappa(\eta+\omega_{m}(4\beta_{1}-2\kappa\gamma_{m}))^{2}]}}{2\eta}\notag\\
+\frac{\sqrt{\kappa[32\beta_{1}\omega_{m}^{2}\eta+\kappa(\eta-\omega_{m}(4\beta_{1}-2\kappa\gamma_{m}))^{2}]}}{2\eta}-\kappa,\quad 
\end{eqnarray}
with $\eta=8\kappa\omega_{m}^{2}-2\beta_{1}\kappa+\gamma_{m}(\kappa^{2}+4\omega_{m}^{2})$. With the parameters above, the width $\Gamma_{\mathrm{OMIT}}\simeq5.998$ which shows an excellent agreement with the numerical result (see the blue-dashed line) in Fig. 2.

\begin{figure}[ptb]
	\includegraphics[width=0.46\textwidth]{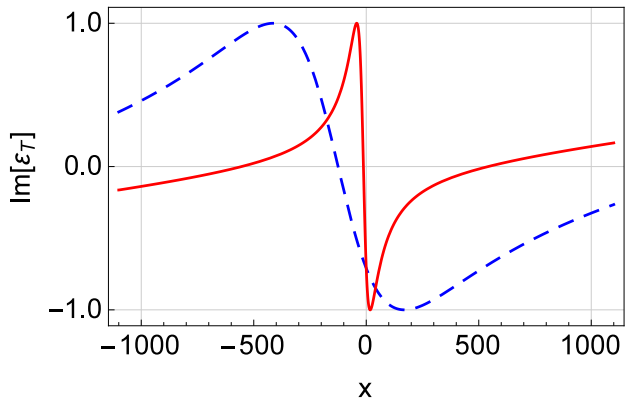}\caption{The imaginary part $\mathrm{Im}[\varepsilon_{T}]$ vs. frequency detuning $x$ with the same parameters as those in Fig. 3.}%
	\label{Fig4}%
\end{figure}

One of the advantages of the perfect OMIT theory is the perfect transparency window can still appear even with a large mechanical damping rate $\gamma_{m}$. In Fig. 3, we plot the  real part  $\mathrm{Re}[\varepsilon_{T}]$ vs. frequency detuning $x$ for large mechanical damping rate $\gamma_{m}=10$ (red-solid) with $\beta_{1}=3\times10^{5}$, and $\beta_{2}=1.25\times10^{4}$ according to Eq. (9), and for $\gamma_{m}=100$ (blue-dashed) with $\beta_{1}=3\times10^{6}$, and $\beta_{2}=1.25\times10^{5}$ according to Eq. (9). The other parameters are $\omega_{m}=\kappa_{1}=\kappa_{2}=10^{4}$. It can be clearly seen from Fig. 3 that the perfect OMIT can do occur with large mechanical damping rate $\gamma_{m}$ as long as the conditions in Eq. (9) are satisfied. In addition, according to Eq. (10), the width $\Gamma_{\mathrm{OMIT}}=59.83$ (red-solid) and $\Gamma_{\mathrm{OMIT}}=583.79$ (blue-dashed), which are consistent with the results in Fig. 3. 
In Fig. 4, we plot the imaginary part $\mathrm{Im}[\varepsilon_{T}]$ vs. frequency detuning $x$ with the same parameters as Fig. 3. It can be seen from Fig. 4 that the dispersion curve becomes very steep at the perfect transparency window and the slope is negative there.

The above discussion is based on the condition that the cavity damping rate $\kappa_{2}$ is much larger than the transparent window position $x$. While if the second cavity is the microwave cavity, the cavity damping rate $\kappa_{2}$ can be very small \cite{Nunnenkamp2014prl,Barzanjeh2012prl}. If $\kappa_{2}$ is very small, the above calculation would be a little more complicated. However, we can always obtain the window position $x$ and driving strength $\beta_{2}$ through numerical methods according to Eq. (7). For example, for parameters $\kappa_{2}=10$, $\omega_{m}=10^{4}$, $\kappa_{1}=4\times10^{3}$, $\gamma_{m}=1$ and $\beta_{1}=10^{5}$, we numerically obtain the window position $x\simeq-5.55$ and $\beta_{2}=5.91$. With these parameters, in Fig. 5 we plot the  real part  $\mathrm{Re}[\varepsilon_{T}]$ vs. frequency detuning $x$, and the inset in Fig. 5 shows a zoom-in of the transparency window at $x\simeq-5.55$.
It can be clearly seen from Fig. 5 that the perfect OMIT can still occur with small cavity damping rate $\kappa_{2}$.

\begin{figure}[ptb]
	\includegraphics[width=0.46\textwidth]{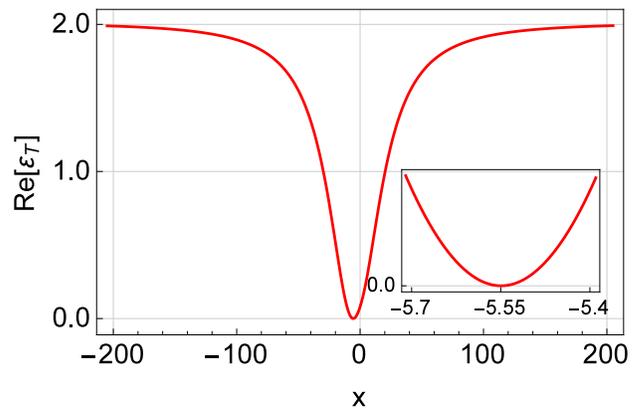}\caption{The  real part  $\mathrm{Re}[\varepsilon_{T}]$ vs. frequency detuning $x$ for $\kappa_{2}=10$ with parameters  $\omega_{m}=10^{4}$, $\kappa_{1}=4\times10^{3}$, $\gamma_{m}=1$, $\beta_{1}=10^{5}$ and $\beta_{2}=5.91$. The inset in Fig. 5 shows a zoom-in of the transparency window at $x\simeq-5.55$.}%
	\label{Fig5}%
\end{figure}

\section{Ultraslow light}

\begin{figure}[ptb]
	\includegraphics[width=0.47\textwidth]{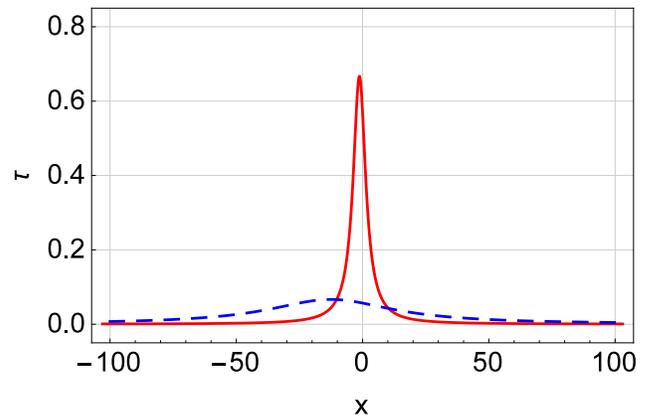}\caption{The time delay $\tau$ vs. frequency detuning $x$ for $\gamma_{m}=1$ (red-solid) with $\beta_{1}=3\times10^{4}$ and $\beta_{2}=1250$ (same as Fig. 2), and for $\gamma_{m}=10$ (blue-dashed) with $\beta_{1}=3\times10^{5}$ and $\beta_{2}=1.25\times10^{4}$ (same as Fig. 3). The other parameters are $\omega_{m}=\kappa_{1}=\kappa_{2}=10^{4}$.}%
	\label{Fig6}%
\end{figure}

\begin{figure}[ptb]
	\includegraphics[width=0.47\textwidth]{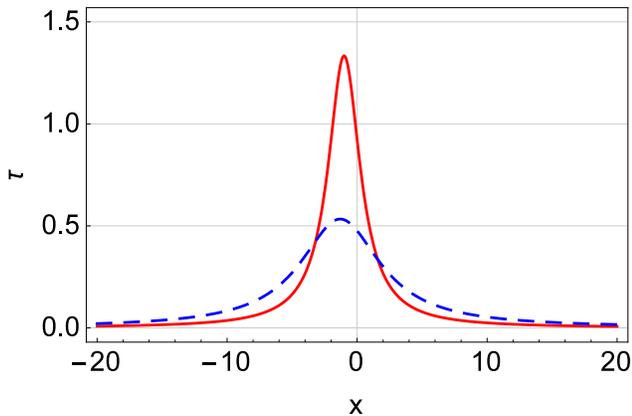}\caption{The time delay $\tau$ vs. frequency detuning $x$ for $\kappa_{1}=\kappa_{2}=2\times10^{4}$ (red-solid) with $\beta_{1}=3\times10^{4}$ and $\beta_{2}=10^{4}$ according to Eq. (9), and for $\kappa_{1}=\kappa_{2}=8\times10^{3}$ (blue-dashed) with $\beta_{1}=3\times10^{4}$ and $\beta_{2}=160$ according to Eq. (9). The other parameters are $\omega_{m}=10^{4}$ and $\gamma_{m}=1$.}%
	\label{Fig7}%
\end{figure}

The time delay (slow light effect) of the probe field with frequency $\omega_{p}$ in the output field can
be determined by \cite{Tarhan2013,Weis2010,Safavi-Naeini2011}
\begin{eqnarray}
\tau=\frac{\partial\mathrm{arg}[\varepsilon_{T}-1]}{\partial\omega_{p}}.
\end{eqnarray}
The positive (negative) value of the time delays represents slow (fast) light \cite{Bigelow2003sci} in the system. 
According to Eqs. (7), (9) and (11), we can obtain the analytic expressions of time delays $\tau$, but it is too tedious to be reported here. However, the time delay at the transparent window can be obtained as   
\begin{eqnarray}
\tau=\frac{\kappa_{1}[8\kappa_{2}\omega_{m}^{2}+\gamma_{m}(\kappa_{2}^{2}+4\omega_{m}^{2})]}{4\beta_{1}\kappa_{2}\omega_{m}^{2}}-\frac{\kappa_{1}^{2}(\kappa_{2}^{2}+4\omega_{m}^{2})}{2\kappa_{2}\omega_{m}^{2}(\kappa_{1}^{2}+4\omega_{m}^{2})}.\quad 
\end{eqnarray}

In Fig. 6, we plot the time delay $\tau$ vs. frequency detuning $x$ for $\gamma_{m}=1$ (red-solid) with $\beta_{1}=3\times10^{4}$ and $\beta_{2}=1250$ (same as Fig. 2), and for $\gamma_{m}=10$ (blue-dashed) with $\beta_{1}=3\times10^{5}$ and $\beta_{2}=1.25\times10^{4}$ (same as Fig. 3). The other parameters are $\omega_{m}=\kappa_{1}=\kappa_{2}=10^{4}$ (same as Fig. 2). It can be seen from Fig. 6 that the time delay $\tau$ exactly takes the maximum at the transparency window where the steepest dispersion appears. In fact, it can be proofed that the time delay at the transparency window is exactly equal to the negative dispersion curve slope there. It means that the steeper the slope of dispersion curve is, the larger the slow light effect becomes. From Fig. 6, the maximum delay $\tau_{\mathrm{max}}\simeq0.67$ for red-solid line and $\tau_{\mathrm{max}}\simeq0.067$ for blue-dashed line, which are very consistent with the results according to Eq. (12).

We also study the effect of cavity damping rate $(\kappa_{1}, \kappa_{2})$ on the time delay $\tau$. For simplicity, we also take $\kappa_{1}=\kappa_{2}$ here. In Fig. 7, we plot the time delay $\tau$ vs. frequency detuning $x$ for $\kappa_{1}=\kappa_{2}=2\times10^{4}$ (red-solid) with $\beta_{1}=3\times10^{4}$ and $\beta_{2}=10^{4}$ according to Eq. (9), and for $\kappa_{1}=\kappa_{2}=8\times10^{3}$ (blue-dashed) with $\beta_{1}=3\times10^{4}$ and $\beta_{2}=160$ according to Eq. (9). The other parameters are $\omega_{m}=10^{4}$ and $\gamma_{m}=1$. 
It can be seen from Fig. 7 that the maximum time delay (at the transparency window) in the unresolved sideband regime ($\kappa_{1}>\omega_{m}$) will larger than that in the  resolved sideband regime ($\kappa_{1}<\omega_{m}$). With the parameters, the maximum time delay $\tau_{\mathrm{max}}\simeq1.33$ in the unresolved sideband regime (see red-solid). It means that the $\tau_{\mathrm{max}}\simeq1.33\,{\rm{sec}}$ (if the units of physical quantities above are Hertz), which is actually a long-lived slow light.
This long-lived slow light may be used for OMIT-based memories in the future.

\section{Optomechanically induced absorption}

\begin{figure}[ptb]
	\includegraphics[width=0.46\textwidth]{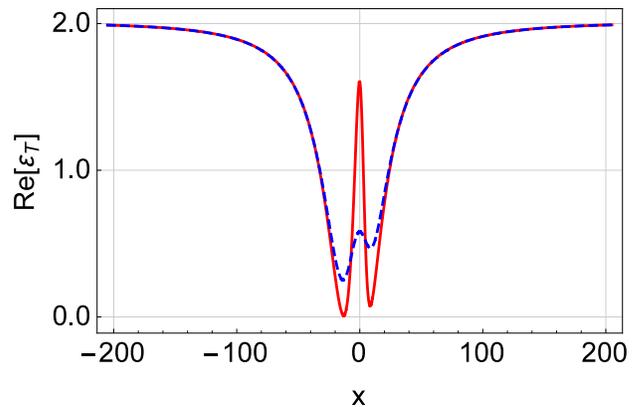}\caption{The real part $\mathrm{Re}[\varepsilon_{T}]$ vs. frequency detuning $x$ for $\kappa_{2}=10$ (blue-dashed) and $\kappa_{2}=1$ (red-solid) with parameters $\kappa_{1}=4\times10^{3}$, $\omega_{m}=10^{4}$, $\gamma_{m}=1$, $\beta_{1}=10^{5}$ and $\beta_{2}=100$.}%
	\label{Fig8}%
\end{figure}

Compared with $\mathrm{Re}[\varepsilon_{T}]=0$ at the window of the perfect OMIT, the phenomenon of optomechanically induced absorption  will appear at the near resonance position $x=0$ ($\mathrm{Re}[\varepsilon_{T}]$ shows a noticeable increase at position $x=0$), if some conditions are satisfied. We first do some qualitative analyses of these conditions. At the position $x=0$, we have
\begin{eqnarray}
\varepsilon_{T}(x=0)=\frac{2\kappa_{1}}{\kappa_{1}+\frac{\beta_{1}}{\frac{\gamma_{m}}{2}-\frac{\beta_{1}}{\kappa_{1}-2i\omega_{m}}+\frac{\beta_{2}}{\kappa_{2}}-\frac{\beta_{2}}{\kappa_{2}-2i\omega_{m}}}},
\end{eqnarray}
which means that $\mathrm{Re}[\varepsilon_{T}]$ will display a noticeable increase at $x=0$ if the ratio $\beta_{2}/\kappa_{2}$ is large enough. In other words, if $\kappa_{2}$ is small enough or the driving strength $\beta_{2}$ is large enough,  the phenomenon of optomechanically induced absorption  at $x=0$ can occur.

In Fig. 8, we plot the $\mathrm{Re}[\varepsilon_{T}]$ vs. the frequency detuning $x$ for $\kappa_{2}=10$ (blue-dashed) and $\kappa_{2}=1$ (red-solid) with parameters $\kappa_{1}=4\times10^{3}$, $\omega_{m}=10^{4}$, $\gamma_{m}=1$, $\beta_{1}=10^{5}$ and $\beta_{2}=100$. It can be seen from Fig. 8 that the phenomenon of optomechanically induced absorption at $x=0$ becomes very significant with the decrease of $\kappa_{2}$. In addition, according to Eq. (13), we have $\varepsilon_{T}(x=0)\simeq0.58$ for $\kappa_{2}=10$ and $\varepsilon_{T}(x=0)\simeq1.60$ for $\kappa_{2}=1$, which are clearly consistent with the numerical results in Fig. 8.

\begin{figure}[ptb]
	\includegraphics[width=0.47\textwidth]{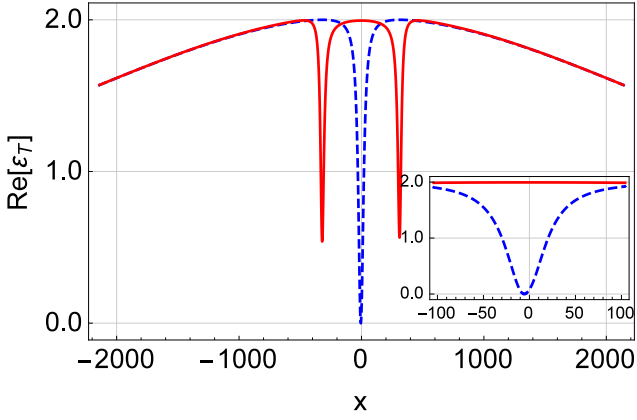}\caption{The real part  $\mathrm{Re}[\varepsilon_{T}]$ vs. frequency detuning $x$ for  $\beta_{2}=5.91$ (blue-dashed) and  $\beta_{2}=10^{5}$ (red-solid) with parameters  $\omega_{m}=10^{4}$, $\kappa_{1}=4\times10^{3}$, $\gamma_{m}=1$, $\beta_{1}=10^{5}$ and $\kappa_{2}=10$.  The inset in Fig. 9 shows a zoom-in of the transparency window at $x\simeq-5.55$.}%
	\label{Fig9}%
\end{figure}

Generally, once a quantum device has been manufactured, its characteristic parameters are fixed. Therefore, it is not easy to achieve optomechanically induced absorption by changing the dissipation rate $\kappa_{2}$. While it is very convenient to adjust the driving strength $\beta_{2}$ in experiments. In Fig. 9, we plot the $\mathrm{Re}[\varepsilon_{T}]$ vs. the frequency detuning $x$ for $\beta_{2}=10^{5}$ (red-solid) with parameters $\kappa_{2}=10$, $\kappa_{1}=4\times10^{3}$, $\omega_{m}=10^{4}$, $\gamma_{m}=1$, and $\beta_{1}=10^{5}$.  With the parameters, $\mathrm{Re}[\varepsilon_{T}]=1.9995$ at $x=0$ according to Eq. (13), which is very much in agreement with the results in Fig. 9. Hence the optomechanically induced absorption can be achieved at $x=0$ with the driving strength $\beta_{2}=10^{5}$. In addition, the absorption curve is approximately a horizontal line over a very wide frequency range, see the red-solid line in the inset in Fig. 9.  For comparison, we also plot the curve of perfect OMIT with the same parameters but $\beta_{2}=5.91$ (blue-dashed), the transparency window $x\simeq-5.55$ (see the blue-dashed line in the inset of Fig. 9). It means that the conversion between perfect OMIT and optomechanically induced absorption of the probe field can be achieved simply by adjusting the driving strength $\beta_{2}$. These results can be used to achieve the optical switch in modern optical networks.

\section{Conclusion}

In summary, we have theoretically studied the controllable optical responses in a two-cavity optomechanical system, especially on the $\mathit{perfect}$ optomechanically induced transparency (OMIT), long-lived slow light and optomechanically induced absorption in the model. From the theoretical results, we can draw some conclusions. First, the perfect OMIT can be still achieved even with a large mechanical damping rate $\gamma_{m}$, which is difficult to be realized in the usual OMIT theory. Secondly, at the transparency window of the perfect OMIT, the long-lived slow light can be achieved, which can be used for OMIT-based memories in the future. Thirdly, an optical switch taking advantage of the conversion between perfect OMIT and optomechanically induced absorption can be achieved just by adjusting the driving field strength of the second cavity. We believe that the results can be used to control optical transmission in quantum information processing.

\appendix{}

\section{Derivation of $a_{1+}$}

Substituting Eq. (4) in Eq. (3), just keeping the constant term and the first order term of $\varepsilon_{p}$, and then comparing the coefficients of the terms $e^{i\delta t}$ and $e^{-i\delta t}$  on both sides of the equation, we can obtain
\begin{eqnarray}
q_{0}=\frac{\hbar g_{0}(|a_{10}|^{2}-|a_{20}|^{2})}{m\omega_{m}^{2}},\\
q_{-}=\frac{\hbar g_{0}(a_{10}a_{1+}^{\ast}+a_{10}^{\ast}a_{1-}-a_{20}a_{2+}^{\ast}-a_{20}^{\ast}a_{2-})}{m(\omega_{m}^{2}-\delta^{2}+i\gamma_{m}\delta)},\\
q_{+}=\frac{\hbar g_{0}(a_{10}^{\ast}a_{1+}+a_{10}a_{1-}^{\ast}-a_{20}^{\ast}a_{2+}-a_{20}a_{2-}^{\ast})}{m(\omega_{m}^{2}-\delta^{2}-i\gamma_{m}\delta)},\\
a_{10}=\frac{\varepsilon_{c}}{\kappa_{1}+i\Delta_{1}},\\
a_{1+}=\frac{ig_{0}q_{+}a_{10}+1}{\kappa_{1}-i(\delta-\Delta_{1})},\\
a_{1-}=\frac{ig_{0}q_{-}a_{10}}{\kappa_{1}+i(\delta+\Delta_{1})},\\
a_{20}=\frac{\varepsilon_{d}}{\kappa_{2}+i\Delta_{2}},\\
a_{2+}=\frac{-ig_{0}q_{+}a_{20}}{\kappa_{2}-i(\delta-\Delta_{2})},\\
a_{2-}=\frac{-ig_{0}q_{-}a_{20}}{\kappa_{2}+i(\delta+\Delta_{2})},
\end{eqnarray}
with $\Delta_{1}=\Delta_{c}-g_{0}q_{0}$ and $\Delta_{2}=\Delta_{d}+g_{0}q_{0}$.

In order to obtain the expression of $a_{1+}$, we need to give the expression of $q_{+}$ in Eq. (A2). To this end, according to Eqs. (A3)--(A9) and using the fact $q_{-}^{\ast}=q_{+}$, we have
\begin{eqnarray}
Aq_{+}=a_{10}^{\ast}a_{1+}+a_{10}a_{1-}^{\ast}-a_{20}^{\ast}a_{2+}-a_{20}a_{2-}^{\ast},\\
a_{10}^{\ast}a_{1+}=B(ig_{0}|a_{10}|^{2}q_{+}+a_{10}^{\ast}),\\
a_{10}a_{1-}^{\ast}=-Cig_{0}|a_{10}|^{2}q_{+},\\
a_{20}^{\ast}a_{2+}=-Dig_{0}|a_{20}|^{2}q_{+},\\
a_{20}a_{2-}^{\ast}=Eig_{0}|a_{20}|^{2}q_{+},
\end{eqnarray}
with
\begin{eqnarray}
A=\frac{m(\omega_{m}^{2}-\delta^{2}-i\gamma_{m}\delta)}{\hbar g_{0}},\\
B=\frac{1}{\kappa_{1}-i(\delta-\Delta_{1})},\\
C=\frac{1}{\kappa_{1}-i(\delta+\Delta_{1})},\\
D=\frac{1}{\kappa_{2}-i(\delta-\Delta_{2})},\\
E=\frac{1}{\kappa_{2}-i(\delta+\Delta_{2})}.
\end{eqnarray}
From Eqs. (A11)--(A14), we can obtain the expression of $a_{10}^{\ast}a_{1+}+a_{10}a_{1-}^{\ast}-a_{20}^{\ast}a_{2+}-a_{20}a_{2-}^{\ast}$, and then combining Eq. (A10), the expression of $q_{+}$ can be obtained as
\begin{eqnarray}
q_{+}=\frac{Ba_{10}^{\ast}}{A-ig_{0}[(B-C)|a_{10}|^{2}+(D-E)|a_{20}|^{2}]}.\quad 
\end{eqnarray}
Substituting Eq. (A20) into Eq. (A5), we obtain $a_{1+}$ as
\begin{eqnarray}
a_{1+}=\frac{ig_{0}B^{2}|a_{10}|^{2}}{A-ig_{0}[(B-C)|a_{10}|^{2}+(D-E)|a_{20}|^{2}]}+B\notag\\
=\frac{[A-ig_{0}(-C|a_{10}|^{2}+(D-E)|a_{20}|^{2})]B}{A-ig_{0}[(B-C)|a_{10}|^{2}+(D-E)|a_{20}|^{2}]}\notag\\
=\frac{1}{\frac{1}{B}-\frac{ig_{0}|a_{10}|^{2}}{A-ig_{0}(-C|a_{10}|^{2}+D|a_{20}|^{2}-E|a_{20}|^{2})}}.\qquad
\end{eqnarray}
Substituting Eqs. (A15)--(A19) into Eq. (A21) and setting $\beta_{1}=\frac{\hbar g_{0}^{2}|a_{10}|^{2}}{2m\omega_{m}}$ and $\beta_{2}=\frac{\hbar g_{0}^{2}|a_{20}|^{2}}{2m\omega_{m}}$, we can obtain the Eq. (5) in text for $a_{1+}$.

\end{document}